\documentclass[12pt,twoside,dvips]{article}
\usepackage{amssymb}
\usepackage{pifont}
\usepackage{amsfonts}
\usepackage{amsmath}
\usepackage[spanish,english]{babel}
\usepackage{graphicx}

\begin{document}

\title{Mass-matrix ansatz and constraints on $B_{s}^{0}-\overline{B}_{s}^{0}$
mixing in 331 models,}
\author{R. Mart\'{\i}nez$\thanks{%
e-mail: remartinezm@unal.edu.co}$ \ and F. Ochoa$\thanks{%
e-mail: faochoap@unal.edu.co}$ \and Departamento de F\'{\i}sica, Universidad
Nacional, \\
%EndAName
Bogot\'{a}-Colombia}
\maketitle

\begin{abstract}
Comparing the theoretically predicted and measured values of 
the mass difference of the $B^{0}_{s}$ system, we estimate the
lower bound on the mass of the $Z^{\prime}$ boson of models based 
on the $SU(3)_{c} \otimes SU(3)_{L} \otimes U(1)_X$ gauge group. By 
assuming zero-texture approaches of the quark mass matrices, 
we find the ratio of the measured value to the theoretical prediction from the Standard Model and the $Z^{\prime}$ contribution from the 331 models of the mass difference of the $B^{0}_{s}$ system. 
We find lower bounds on the $Z^{\prime}$ mass ranging between $1$ TeV and $30$ TeV for 
the two most popular 331 models, and four different zero-textures 
ans\"atze. The above results
are expressed as a function of the weak angle associated to the 
$b-s-Z^{\prime}$ couplings.

\end{abstract}

\section{Introduction}

Although the Standard Model (SM) \cite{ME} is considered as an effective low energy
theory that should be embedded into a more fundamental theory, many of the
SM predictions have been successfully tested by precision
measurements. The latter impose strong restrictions to new physics
contributions associated to any extension of the SM \cite{precision}. Thus, small
deviations between the experimental data and the SM predictions allow to
set stringent limits on new physics from a more fundamental theory that
contains new types of matter and interactions at the TeV scale. It 
will be explored with the new generation of accelerators and detectors like the
forthcoming Large Hadron Collider (LHC) \cite{LHC}. Among the possible extensions of the SM, the
models with gauge symmetry $SU(3)_{c} \otimes SU(3)_{L} \otimes U(1)_{X}$,
also called 3-3-1 models \cite{ten,frampton}, arise as an interesting alternative with new
physics content and some motivating features. First of all, from the cancellation
of chiral anomalies \cite{anomalias} and asymptotic freedom in QCD, the 3-3-1 models
can explain why there are three fermion families. Secondly, since the
third family is treated under a different representation, the large mass
difference between the heaviest quark family and the two lighter ones may be
understood \cite%
{third-family}. Thirdly, the models have a scalar content similar to the two Higgs
doublet model (2HDM), which allow to predict the quantization of electric
charge and the vectorial character of the electromagnetic interactions \cite{quantum-charge,vectorlike}.
Also, these models contain a natural Peccei-Quinn symmetry , necessary to
solve the strong-CP problem \cite{PC,PC331}. Finally, the model introduces new types of
matter relevant to the next generations of colliders at the TeV energy
scales, which do not spoil the low energy limits at the electroweak scale.

In the SM, the Flavor Changing Neutral Currents (FCNC) are strongly suppresed
with respect to the charged-current weak interactions, which follows from
the experimental data on neutral meson decays and the mass difference in
meson systems exhibiting particle-antiparticle mixing \cite{Wu}. In particular, some extensions of the SM produce new FCNC contributions at tree level, as for example some models with an extra neutral $Z^{\prime}$ boson, which represents an stringent limit for new physics. Although not all models with new neutral $Z^{\prime}$ bosons exhibit additional FCNC contributions \cite{david}, many interesting ones contain FCNC effects at tree level \cite{fcnc-zprimas,mesones,mesones2}. In 3-3-1
models, the contributions in meson systems have been considered before \cite{dumm-pleitez} in $K^{0}-\overline{K^{0}},$ and $%
B_{d}^{0}-\overline{B_{d}^{0}}$ systems, which induce the flavor changing
transitions $s \leftrightarrow d$ and $b \leftrightarrow d$, respectively,
while no information other than a lower bound associated to the $b
\leftrightarrow s$ transition was available for the $B_{s}^{0}-\overline{%
B_{s}^{0}}$ system. However, the $b-s$ sector was recently confirmed in the $%
B_{s}$ mixing by both CDF and DO\hspace{-0.3cm}/ \cite{data-particle}:

\begin{eqnarray*}
\text{CDF }\text{: } &&\Delta M_{s}=17.33_{-0.21}^{+0.42}\text{ ps}^{-1}, \\
\text{DO\hspace{-0.3cm}/ }\text{: } &&\Delta M_{s}=19.0\pm 1.215\text{ ps}%
^{-1}
\end{eqnarray*}

\noindent Ref. \cite{mesones2} use the following averages

\begin{eqnarray}
&&\Delta M_{s}^{exp}=17.46_{-0.3}^{+0.47}\text{ ps}^{-1},  \notag \\
&&\Delta M_{s}^{SM}=19.52\pm 5.28\text{ ps}^{-1},  \label{massdiff}
\end{eqnarray}

\noindent for the experimental and SM prediction, respectively. Since the study of $B$ physics has been an important tool to extract information on CP violation and new physics \cite{bs}, we will use the above data for the mass difference of the $%
B_{s}$ system to explore the FCNC contribution induced by the $Z^{\prime }$ boson in the two most popular 3-3-1 models. However, since FCNC contribution in these models are very sensitive to the
rotations of the fermionic spectrum to mass eigenstates, it is neccesary to
implement some criterion to fix the values of the components of the
rotation matrices, and to get numerical predictions on the
meson mass diference. In contrast to other studies in $%
D^{0}$, $K^{0}$ and $%
B_{d}^{0}$ systems \cite{dumm-pleitez}, we will consider various cases for the rotation matrix, including the texture-zero approches, where an
ansatz on the texture of the femion mass matrices is adopted in agreement
with the measured masses and mixing angles of the Cabibbo-Kobayashi-Maskawa matrix. An additional motivation to study the $%
B_{s}$ system comes from the fact that the $b-s$ sector induces the maximum flavor-changing contribution, as will be confirmed in this work. This offers a good opportunity to extract information on new physics at low energy.

Eq. (\ref{massdiff}) shows good agreement between the experimental data and the SM one-loop prediction of $\Delta M_{s}$, however, due to the hadronic parameters, the SM prediction contains a large uncertainty which we use to find allowed regions for the mass of the $Z^{\prime }$
boson and the weak angle associated to the $b-s-Z^{\prime}$ coupling by assuming four specific forms in the rotation matrix of the quark mass.

\section{The 331 spectrum}

The fermionic structure is shown in Tab. \ref{tab:espectro} where all
leptons transform as $(\mathbf{3,X}_{\ell }^{L})$ and $(\mathbf{1,X}_{\ell
}^{R})$ under the $\left(SU(3)_{L},U(1)_{X}\right) $ sector, with $\mathbf{X}%
_{\ell }^{L}$ and $(\mathbf{X}_{\ell }^{R})$ the $U(1)_{X}$ generators
associated with the left- and right-handed leptons, respectively; while the
quarks transform as $(\mathbf{3}^{\ast }\mathbf{,X}_{q_{m^{\ast }}}^{L})$, $%
(\mathbf{1,X}_{q_{m^{\ast }}}^{R})$ for the first two families, and $(%
\mathbf{3,X}_{q_{3}}^{L})$, $(\mathbf{1,X}_{q_{3}}^{R})$ for the third
family, each one with its $U(1)_{X}$ values for the left- and right-handed
quarks. The quantum numbers $\mathbf{X}_{\psi }$ for each representation are
given in the third column from Tab. \ref{tab:espectro}, where the electric
charge is defined by

\begin{equation}
Q=T_{3}+\beta T_{8}+XI,  \label{charge}
\end{equation}

\noindent with $T_{3} = 1/2$diag$(1,-1,0)$, $T_{8}=(1/2\sqrt{3})$diag$(1,1,-2)$ and $%
\beta=-1/\sqrt{3}$ and $-\sqrt{3}$, where the first case contains the
Foot-Long-Truan model (FLT) \cite{twelve} and the second contains the
Pisano-Pleitez-Frampton model (PPF) \cite{ten,frampton}.

\begin{table}[tbp]
\begin{center}
\begin{equation*}
\begin{tabular}{c||c||c}
\hline\hline
$representation$ & $Q_{\psi }$ & $X_{\psi }$ \\ \hline\hline
$\ 
\begin{tabular}{c}
$q_{m^{\ast }L}=\left( 
\begin{array}{c}
d_{m^{\ast }} \\ 
-u_{m^{\ast }} \\ 
J_{m^{\ast }}%
\end{array}%
\right) _{L}\mathbf{3}^{\ast }$ \\ 
\\ 
\\ 
$d_{m^{\ast }R};$ $u_{m^{\ast }R};$ $J_{m^{\ast }R}:\mathbf{1}$%
\end{tabular}%
\ $ & 
\begin{tabular}{c}
$\left( 
\begin{array}{c}
-\frac{1}{3} \\ 
\frac{2}{3} \\ 
\frac{1}{6}+\frac{\sqrt{3}\beta }{2}%
\end{array}%
\right) $ \\ 
\\ 
$-\frac{1}{3};$ $\frac{2}{3};$ $\frac{1}{6}+\frac{\sqrt{3}}{2}\beta $%
\end{tabular}
& 
\begin{tabular}{c}
\\ 
$X_{q_{m^{\ast }}}^{L}=\frac{1}{6}+\frac{\beta }{2\sqrt{3}}$ \\ 
\\ 
\\ 
$X_{d_{m^{\ast }},u_{m^{\ast }},J_{m^{\ast }}}^{R}=-\frac{1}{3},\frac{2}{3},%
\frac{1}{6}+\frac{\sqrt{3}}{2}\beta $%
\end{tabular}
\\ \hline\hline
\begin{tabular}{c}
$q_{3L}=\left( 
\begin{array}{c}
u_{3} \\ 
d_{3} \\ 
J_{3}%
\end{array}%
\right) _{L}:\mathbf{3}$ \\ 
\\ 
$u_{3R};$ $d_{3R};$ $J_{3R}:\mathbf{1}$%
\end{tabular}
& 
\begin{tabular}{c}
$\left( 
\begin{array}{c}
\frac{2}{3} \\ 
-\frac{1}{3} \\ 
\frac{1}{6}-\frac{\sqrt{3}\beta }{2}%
\end{array}%
\right) $ \\ 
\\ 
$\frac{2}{3};$ $-\frac{1}{3};$ $\frac{1}{6}-\frac{\sqrt{3}\beta }{2}$%
\end{tabular}
& 
\begin{tabular}{c}
\\ 
$X_{q^{(3)}}^{L}=\frac{1}{6}-\frac{\beta }{2\sqrt{3}}$ \\ 
\\ 
\\ 
$X_{u_{3},d_{3},J_{3}}^{R}=\frac{2}{3},-\frac{1}{3},\frac{1}{6}-\frac{\sqrt{3%
}\beta }{2}$%
\end{tabular}
\\ \hline\hline
\begin{tabular}{c}
$\ell _{jL}=\left( 
\begin{array}{c}
\nu _{j} \\ 
e_{j} \\ 
E_{j}^{-Q_{1}}%
\end{array}%
\right) _{L}:\mathbf{3}$ \\ 
\\ 
$e_{jR};$ $E_{jR}^{-Q_{1}}$%
\end{tabular}
& 
\begin{tabular}{c}
$\left( 
\begin{array}{c}
0 \\ 
-1 \\ 
-\frac{1}{2}-\frac{\sqrt{3}\beta }{2}%
\end{array}%
\right) $ \\ 
\\ 
$-1;$ $-\frac{1}{2}-\frac{\sqrt{3}\beta }{2}$%
\end{tabular}
& 
\begin{tabular}{c}
\\ 
$X_{\ell _{j}}^{L}=-\frac{1}{2}-\frac{\beta }{2\sqrt{3}}$ \\ 
\\ 
\\ 
$X_{e_{j},E_{j}}^{R}=-1,$ $-\frac{1}{2}-\frac{\sqrt{3}\beta }{2}$%
\end{tabular}
\\ \hline\hline
\end{tabular}%
\end{equation*}%
\end{center}
\caption{\textit{Fermionic content for three generations with\ }$\protect%
\beta = -1/\sqrt{3}, -\sqrt{3}\ $\textit{. We take} $m^{\ast }=1,2$ \textit{and} $j=1,2,3$}
\label{tab:espectro}
\end{table}

For the scalar sector, we introduce the triplet field $\chi $ with vacuum
expectation value (VEV) $\left\langle \chi \right\rangle ^{T}=\left( 0,0,\nu
_{\chi }\right) $, which provides the masses of the third fermionic
components. In the second transition, it is necessary to introduce two
triplets$\;\rho $ and $\eta $ with VEV $\left\langle \rho \right\rangle
^{T}=\left( 0,\nu _{\rho },0\right) $ and $\left\langle \eta \right\rangle
^{T}=\left( \nu _{\eta },0,0\right) $, in order to give masses to the quarks
of up- and down-type, respectively \cite{331us}.

In the gauge boson spectrum associated with the group $SU(3)_{L}\otimes
U(1)_{X},$ we are just interested in the physical neutral sector that
corresponds to the photon, $Z$, and $Z^{\prime },$ which are written in
terms of the electroweak basis for $\beta=-1/\sqrt{3}$ and $-\sqrt{3}$ as \cite{beta-arbitrary}

\begin{eqnarray}
A_{\mu } &=&S_{W}W_{\mu }^{3}+C_{W}\left( \beta T_{W}W_{\mu }^{8}+\sqrt{%
1-\beta ^{2}T_{W}^{2}}B_{\mu }\right) ,  \notag \\
Z_{\mu } &=&C_{W}W_{\mu }^{3}-S_{W}\left( \beta T_{W}W_{\mu }^{8}+\sqrt{%
1-\beta ^{2}T_{W}^{2}}B_{\mu }\right) ,  \notag \\
Z_{\mu }^{\prime } &=&-\sqrt{1-\beta ^{2}T_{W}^{2}}W_{\mu }^{8}+\beta
T_{W}B_{\mu },
\end{eqnarray}

\noindent where the Weinberg angle is defined as \cite{beta-arbitrary}

\begin{equation}
S_{W}=\sin \theta _{W}=\frac{g_{X}}{\sqrt{g_{L}^{2}+\left( 1+\beta
^{2}\right) g_{X}^{2}}}
\end{equation}

\noindent and $g_{L},$ $g_{X}$ correspond to the coupling constants of the
groups $SU(3)_{L}$ and $U(1)_{X}$, respectively.

\section{Neutral Couplings}

Using the fermionic content in weak eigenstates from Tab. \ref{tab:espectro}%
, we obtain the neutral coupling for the SM quarks \cite{beta-arbitrary}

\begin{equation}
\mathcal{L}_{D}^{NC}=\frac{g_{L}}{2C_{W}}\left[ \overline{Q^{0}}\gamma _{\mu
}\left( g_{v}^{Q^{0}}-g_{a}^{Q^{0}}\gamma _{5}\right) Q^{0}Z^{\mu }+%
\overline{Q^{0}}\gamma _{\mu }\left( \widetilde{g}_{v}^{Q^{0}}-\widetilde{g}%
_{a}^{Q^{0}}\gamma _{5}\right) Q^{0}Z^{\mu \prime }\right] ,
\label{L-neutro}
\end{equation}

\noindent where $Q^{0}:U^{0}=(u,c,t)^{0}, D^{0}=(d,s,b)^{0}$ for up- and down-type
quarks, respectively. The vector and axial-vector couplings of the $Z$ boson
are

\begin{eqnarray}
g_{v}^{U^{0}} &=&\frac{1}{2}-2Q_{U^{0}}S_{W}^{2},\qquad \qquad g_{a}^{U^{0}}=%
\frac{1}{2},  \notag \\
g_{v}^{D^{0}} &=&-\frac{1}{2}-2Q_{D^{0}}S_{W}^{2},\qquad \quad
\;g_{a}^{U^{0}}=-\frac{1}{2},  \label{Z-couplings}
\end{eqnarray}

\noindent with $Q_{U^{0},D^{0}}$ the electric charge of each quark given by Tab. \ref%
{tab:espectro}; while the corresponding couplings to $Z^{\prime }$ are given
by

\begin{eqnarray}
\widetilde{g}_{v,a}^{U^{0}} &=&\frac{g_{X}C_{W}}{2g_{L}T_{W}}\left[ \frac{1}{%
\sqrt{3}}\left( diag\left( 1,1,-1\right) +\frac{\beta T_{W}^{2}}{\sqrt{3}}%
\right) \pm 2Q_{U^{0}}\beta T_{W}^{2}\right] ,  \notag \\
\widetilde{g}_{v,a}^{D^{0}} &=&\frac{g_{X}C_{W}}{2g_{L}T_{W}}\left[ \frac{1}{%
\sqrt{3}}\left( diag\left( 1,1,-1\right) +\frac{\beta T_{W}^{2}}{\sqrt{3}}%
\right) \pm 2Q_{D^{0}}\beta T_{W}^{2}\right] ,  \label{Zprima-couplings}
\end{eqnarray}

\noindent which are written for $\beta=-1/\sqrt{3}$ and $-\sqrt{3}$. In particular, for the $%
Z^{\prime}$ coupling in the neutral Lagrangian in Eq. (\ref{L-neutro}), we can
write

\begin{equation}
\mathcal{L}^{Z^{\prime }}=\frac{g_{L}}{2C_{W}}\left[ \overline{Q^{0}}\gamma
^{\mu }\left( \widetilde{\epsilon }_{L}^{Q^{0}}P_{L}+\widetilde{\epsilon }%
_{R}^{Q^{0}}P_{R}\right) Q^{0}Z_{\mu }^{\prime }\right] ,  \label{L-neutro2}
\end{equation}

\noindent where $\widetilde{\epsilon }_{L,R}^{Q^{0}}=(1/2)(\widetilde{g}%
_{v}^{Q^{0}}\pm \widetilde{g}_{a}^{Q^{0}}),$ and $P_{L,R}=(1/2)(1\mp \gamma
_{5})$ the chiral projectors. Using the neutral $Z^{\prime}$-couplings from
Eq. (\ref{Zprima-couplings}), the new chiral couplings $\widetilde{\epsilon }%
_{L,R}^{U^{0},D^{0}}$ are written as follows

\begin{eqnarray}
\widetilde{\epsilon }_{L}^{U^{0},D^{0}} &=&\frac{g_{X}C_{W}}{2g_{L}T_{W}}%
\left[ \frac{1}{\sqrt{3}}diag(1,1,-1)+\frac{1}{3}\beta T_{W}^{2}\right] , 
\notag \\
\widetilde{\epsilon }_{R}^{U^{0},D^{0}} &=&\frac{g_{X}C_{W}}{g_{L}T_{W}}%
\left[ Q_{U^{0},D^{0}}\beta T_{W}^{2}\right] .  \label{acoples-quirales}
\end{eqnarray}

On the other hand, we will consider linear combinations among the three
families of quarks to obtain couplings in mass eigenstates 

\begin{equation}
Q^{0}=R_{Q}Q,  \label{mass-rot}
\end{equation}

\noindent where $Q:U=(u,c,t), D=(d,s,b)$ denotes the quarks in mass
eigensates, $Q^{0}$ in weak eigensates and $R_{Q}$ the rotation matrix that
diagonalize the Yukawa mass terms. Thus, we can write the Eq. (\ref{L-neutro2})
as

\begin{equation}
\mathcal{L}^{Z^{\prime }}=\frac{g_{L}}{2C_{W}}\left[ \overline{Q}\gamma
^{\mu }\left( \widetilde{B}_{L}^{Q}P_{L}+\widetilde{B}_{R}^{Q}P_{R}\right)
QZ_{\mu }^{\prime }\right] ,  \label{L-mass}
\end{equation}

\noindent where the chiral couplings in mass eigenstates are defined as

\begin{equation}
\widetilde{B}_{L,R}^{Q}=R_{Q}^{\dag }\widetilde{\epsilon }%
_{L,R}^{Q^{0}}R_{Q}.  \label{chiral-mass}
\end{equation}

\noindent Because of the fact that $\widetilde{\epsilon }_{R}^{Q^{0}}$ in Eq. (\ref%
{acoples-quirales}) is family independent, the right-handed couplings remain
flavor-diagonal in the mass eigenbasis, such that $\widetilde{B}_{R}^{Q}=%
\widetilde{\epsilon }_{R}^{Q^{0}}$. However, due to the $diag(1,1,-1)$ term
from Eq. (\ref{acoples-quirales}) (family dependent couplings), we obtain
non-diagonal components in the left-handed couplings $\widetilde{B}_{L}^{Q}$
in Eq. (\ref{chiral-mass}), which is sensitive to the form of the rotation
matrix $R_{Q}$. In order to have a predictive model, we adopt a different ansatz on the texture of the
quark mass matrices in agreement with the six quark physical masses and the
four physical parameters of the CKM matrix. The $%
SU(3)_{L}\otimes U(1)_{X}$ Lagrangian for the Yukawa interaction between
quarks is

\begin{eqnarray}
-\mathcal{L}_{Yuk} &=&\sum_{m=1}^{2}\overline{q_{m^{\ast }L}}\left[ \Gamma
_{\eta }^{m^{\ast }D}\eta D_{R}^{0}+\Gamma _{\rho }^{m^{\ast }U}\rho
U_{R}^{0}+\Gamma _{\chi }^{m^{\ast }J}\chi J_{m^{\ast }R}^{0}\right]  \notag
\\
&&+\overline{q_{3L}}\left[ \Gamma _{\rho }^{3D}\rho D_{R}^{0}+\Gamma _{\eta
}^{3U}\eta U_{R}^{0}+\Gamma _{\chi }^{3J}\chi J_{3R}^{0}\right] +h.c,
\label{yukawa-1}
\end{eqnarray}%
with $\eta$ and $\rho $ being the two scalar triplets neccesary to give masses to the SM fermion spectrum from Table \ref%
{tab:espectro}, and $\chi $ the scalar triplet that gives masses to the new extra fermions $J_{1,2,3}, E_{1,2,3}$, as explained in Sec. 2. Thus, we are not interested in the couplings  of $\chi$. $\Gamma _{\phi }^{iQ}$ are the Yukawa interaction
matrices. Taking into account only the $SU(2)_{L}$ sector (which lies in
the two upper components of each scalar triplet), and omitting the couplings of $%
\chi $, the mass eigenstates of the scalar sector can be written as \cite%
{beta-arbitrary} 
\begin{eqnarray}
H &=&\left( 
\begin{array}{c}
\phi _{1}^{\mp } \\ 
h_{3}^{0}+\nu \mp i\phi _{3}^{0}%
\end{array}%
\right) =\rho S_{\beta }-\eta ^{\ast }C_{\beta },  \notag \\
\phi &=&\left( 
\begin{array}{c}
h_{2}^{\mp } \\ 
-h_{4}^{0}\mp ih_{1}^{0}%
\end{array}%
\right) =\rho C_{\beta }+\eta ^{\ast }S_{\beta },  \label{scalar-rotation}
\end{eqnarray}%
where $\eta ^{\ast }$ denotes the conjugate representation of $\eta ,$ $\tan
\beta =\nu _{\rho }/\nu _{\eta }$ and $\nu =\sqrt{\nu _{\rho }^{2}+\nu
_{\eta }^{2}}.$ Thus, after some algebraic manipulation, the neutral
couplings of the Yukawa Lagrangian can be written as

\begin{eqnarray}
-\mathcal{L}_{Yuk}^{(0)} &=&\left[ \overline{D_{L}^{0}}\left( M_{D^{0}}\right)
D_{R}^{0}+\overline{U_{L}^{0}}\left( M_{U^{0}}\right) U_{R}^{0}\right] \left( 1+%
\frac{h_{3}^{0}\mp i\phi _{3}^{0}}{\nu }\right)  \notag \\
&&+\left[ \overline{D_{L}^{0}}\left( \Gamma _{D^{0}}\right) D_{R}^{0}+\overline{%
U_{L}^{0}}\left( \Gamma _{U^{0}}\right) U_{R}^{0}\right] \left( h_{4}^{0}\pm
ih_{1}^{0}\right) +h.c,  \label{yukawa-2}
\end{eqnarray}%
where the fermion masses and Yukawa coupling matrices are given by

\begin{equation}
M_{q^{0}}=\nu \left( \Gamma _{1}C_{\beta }+\Gamma _{2}S_{\beta }\right) \text{ \quad
and\quad\ }\Gamma _{q^{0}} =\Gamma _{1}S_{\beta }-\Gamma _{2}C_{\beta },
\label{yukawa-matrix}
\end{equation}%
where $\Gamma _{1}=\Gamma _{\eta }$ and $\Gamma _{2}=\Gamma _{\rho }$. The
Lagrangian from Eq. (\ref{yukawa-2}) is equivalent to the two-Higgs-doublet
model (2HDM) Lagrangian \cite{zhou}, which exhibits FCNC due to the non-diagonal
components of $\Gamma .$ In the literature, there are
various approches on the zero-textures of the quark mass matrices $M_{q^{0}}$ from Eq. (\ref{yukawa-matrix}), where the most popular are listed as follows

\vspace{0.3cm}

\ding{172} \underline{\textbf{Fritzsch ansatz}}: In the basis $%
U^{0}(D^{0})=(u^{0}(d^{0}),c^{0}(s^{0}),t^{0}(b^{0}))$ the quark mass matrices in the Fritzsch ansatz are defined as \cite{frit}

\begin{equation}
\quad \widehat{M}_{q^{0}}=\left( 
\begin{array}{ccc}
0 & \left\vert D_{q}\right\vert & 0 \\ 
\left\vert D_{q}\right\vert & 0 & \left\vert F_{q}\right\vert \\ 
0 & \left\vert F_{q}\right\vert & \left\vert C_{q}\right\vert%
\end{array}%
\right) ,  \label{texture-Fritzsch}
\end{equation}

\noindent with $\left\vert C_{q}\right\vert \approx m_{t,b},$ $\left\vert
F_{q}\right\vert \approx \sqrt{m_{t,b}m_{c,s}}$ and $\left\vert
D_{q}\right\vert \approx \sqrt{m_{u,d}m_{c,s}},$ where $m_{q}$ corresponds
to the physical mass of the quarks. The above ansatz is diagonalized by the
following rotation matrices for both the up- and down-type quarks

\begin{equation}
\quad R_{q}=\left( 
\begin{array}{ccc}
1 & \sqrt{\frac{m_{u,d}}{m_{c,s}}} & -\sqrt{\frac{m_{u,d}}{m_{t,b}}} \\ 
-\sqrt{\frac{m_{u,d}}{m_{c,s}}} & 1 & -\sqrt{\frac{m_{c,s}}{m_{t,b}}} \\ 
0 & \sqrt{\frac{m_{c,s}}{m_{t,b}}} & 1%
\end{array}%
\right) .  \label{Fritzsch-biunitary}
\end{equation}

\vspace{0.3cm}

\ding{173} \underline{\textbf{Matsuda-Nishihura ansatz}}: This texture
takes the same form as Eq. (\ref{texture-Fritzsch}), but with $\left\vert
B_{q}\right\vert =m_{c,s}$ in the (2,2) component of the mass matrix \cite%
{4zeros}. This form have the following rotation matrices

\begin{equation}
\quad R_{q}=\left( 
\begin{array}{ccc}
1 & \sqrt{\frac{m_{u,d}}{m_{c,s}}} & \sqrt{\frac{m_{c,s}m_{u,d}^{2}}{%
m_{t,b}^{3}}} \\ 
-\sqrt{\frac{m_{u.d}}{m_{c,s}}} & 1 & \sqrt{\frac{m_{u,d}}{m_{t,b}}} \\ 
\sqrt{\frac{m_{u,d}^{2}}{m_{c,s}m_{t,b}}} & -\sqrt{\frac{m_{u,d}}{m_{t,b}}}
& 1%
\end{array}%
\right) .  \label{XIng-biunitary}
\end{equation}

\noindent The above ansatz was reconsidered by the authors in ref. \cite%
{fitmatsuda}, where the parameter $C_{q}$ in the $(3,3)$ component is taken as a free
parameter. In particular, they define the ratio $x_{q}=C_{q}/m_{t,b}$, such that the experimental
values of the CKM matrix are derived by fine tuning of the parameter $x_{q}$. Thus, the non-zero components of the mass matrix
takes the form $\left\vert D_{q}\right\vert =\sqrt{m_{c,s}m_{u,d}/x_{q}}$ in
the $(1,2)$ components, $\left\vert B_{q}\right\vert
=m_{t,b}(1-x_{q})+m_{c,s}-m_{u,d}$ in the $(2,2)$ component, and $\left\vert
F_{q}\right\vert =\sqrt{%
(m_{t,b}x_{q}+m_{u,d})(m_{t,b}x_{q}-m_{c,s})(1-x_{q})/x_{q}}$ in the $(2,3)$
components, where the hierarchy $m_{c,s}\ll C_{q}<m_{t,b}$ is
required. The rotation matrix is

\begin{equation}
\quad R_{q}=\left( 
\begin{array}{ccc}
1 & \sqrt{\frac{m_{u,d}}{m_{c,s}}} & \sqrt{\frac{m_{c,s}m_{u,d}(1-x_{q})}{%
m_{t,b}^{2}x_{q}}} \\ 
-\sqrt{\frac{m_{u.d}x_{q}}{m_{c,s}}} & \sqrt{x_{q}} & \sqrt{1-x_{q}} \\ 
\sqrt{\frac{m_{u,d}}{m_{c,s}}(1-x_{q})} & -\sqrt{1-x_{q}} & \sqrt{x_{q}}%
\end{array}%
\right) .  \label{Matsuda2-Biunitary}
\end{equation}

\noindent The authors in ref. \cite%
{fitmatsuda} obtain the values $x_{u}=0.9560$ and $x_{d}=0.9477$.

\vspace{0.3cm}

\ding{174} \underline{\textbf{Matsuda ansatz}}: Another consistent
possibility is to consider different texture assignment for the up- and down-type
quarks, as follows \cite{2matsuda,matsuda}

\begin{equation}
\quad \widehat{M}_{q^{0}}=\left( 
\begin{array}{ccc}
0 & \left\vert D_{q}\right\vert & \left\vert D_{q}\right\vert \\ 
\left\vert D_{q}\right\vert & \left\vert B_{q}\right\vert & \left\vert
F_{q}\right\vert \\ 
\left\vert D_{q}\right\vert & \left\vert F_{q}\right\vert & \left\vert
B_{q}\right\vert%
\end{array}%
\right) ,  \label{texture-Matsuda}
\end{equation}

\noindent with $\left\vert B_{U}\right\vert =(m_{t}+m_{c}-m_{u})/2,$ $\left\vert
F_{U}\right\vert =(m_{t}-m_{c}-m_{u})/2$ and $\left\vert D_{U}\right\vert
\approx \sqrt{m_{t}m_{u}/2}$ for the up sector, while for the down sector the
structure is $\left\vert B_{D}\right\vert =(m_{b}+m_{s}-m_{d})/2,$ $%
\left\vert F_{D}\right\vert =(m_{s}-m_{b}-m_{d})/2$ and $\left\vert
D_{D}\right\vert \approx \sqrt{m_{s}m_{d}/2}.$ The above textures are
diagonalized by \cite{matsuda}

\begin{equation}
\text{ }R_{U}=\left( 
\begin{array}{ccc}
c^{\prime } & 0 & s^{\prime } \\ 
-\frac{s^{\prime }}{\sqrt{2}} & -\frac{1}{\sqrt{2}} & \frac{c^{\prime }}{%
\sqrt{2}} \\ 
-\frac{s^{\prime }}{\sqrt{2}} & \frac{1}{\sqrt{2}} & \frac{c^{\prime }}{%
\sqrt{2}}%
\end{array}%
\right) ;\quad R_{D}=\left( 
\begin{array}{ccc}
c & s & 0 \\ 
-\frac{s}{\sqrt{2}} & \frac{c}{\sqrt{2}} & -\frac{1}{\sqrt{2}} \\ 
-\frac{s}{\sqrt{2}} & \frac{c}{\sqrt{2}} & \frac{1}{\sqrt{2}}%
\end{array}%
\right) ,  \label{Matsuda-Biunitary}
\end{equation}

\noindent where

\begin{eqnarray}
c &=&\sqrt{\frac{m_{s}}{m_{d}+m_{s}}};\qquad s=\sqrt{\frac{m_{d}}{m_{d}+m_{s}%
}};  \notag \\
c^{\prime } &=&\sqrt{\frac{m_{t}}{m_{t}+m_{u}}};\qquad s^{\prime }=\sqrt{%
\frac{m_{u}}{m_{t}+m_{u}}}.  \label{rot-coef}
\end{eqnarray}

\section{$B_{s}^{0}-\overline{B_{s}^{0}}$ mixing constraints}

\begin{figure}[t]
\centering \includegraphics[scale=0.56]{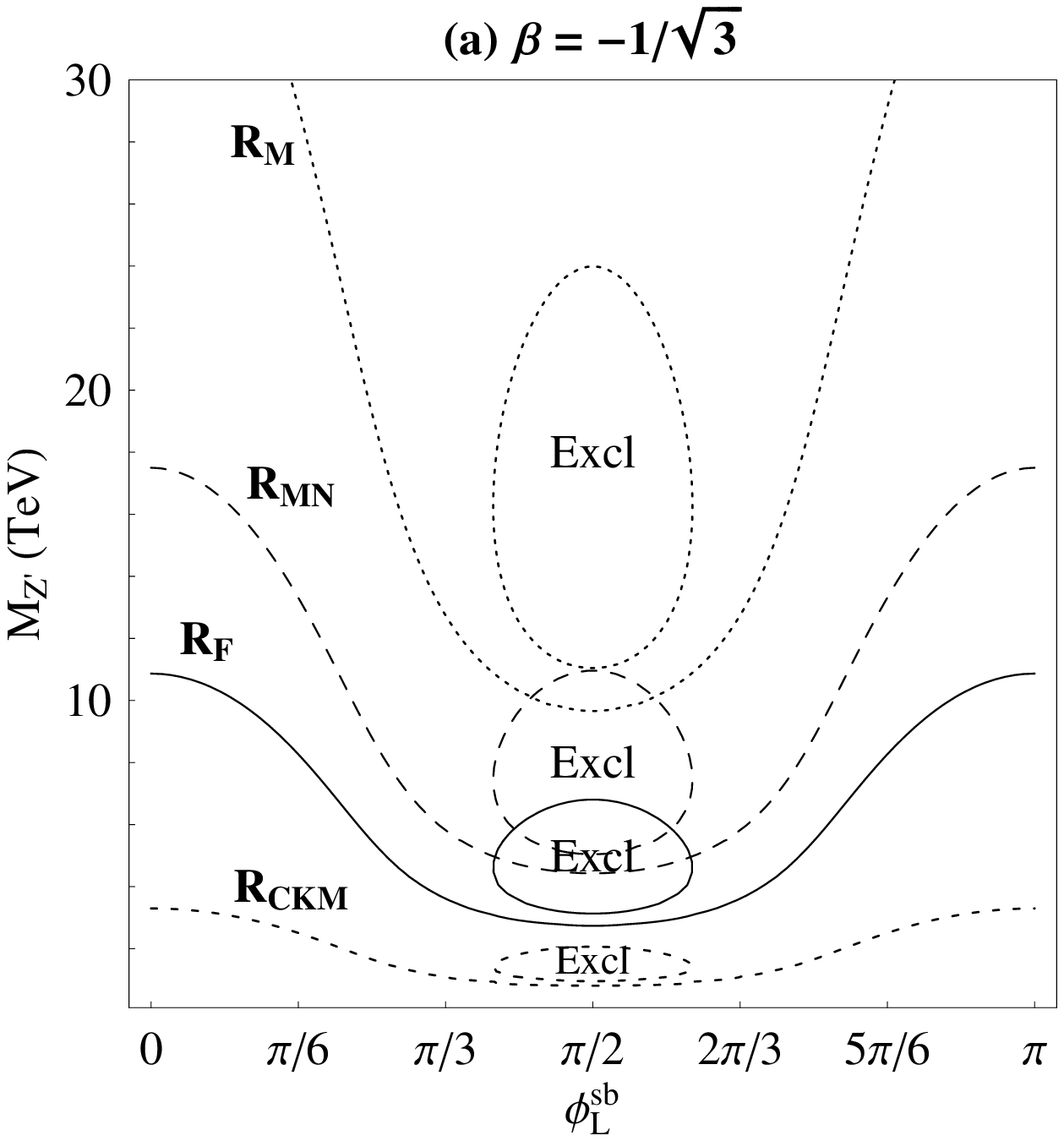} %
\includegraphics[scale=0.58]{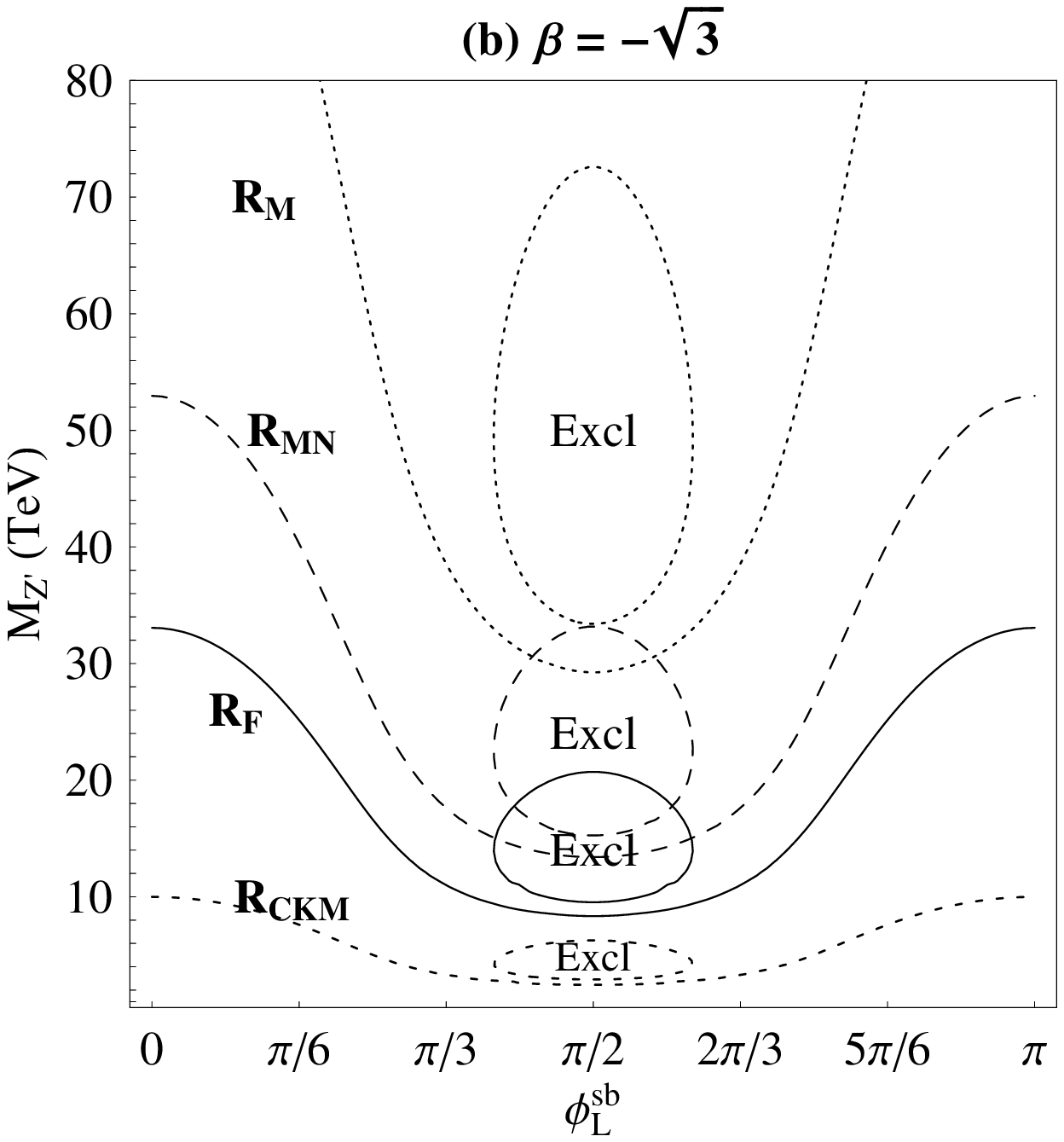}
\caption{\textsf{\ {\protect\small Allowed regions at 1}}$\protect\sigma $
C.L. \textsf{{\protect\small of the $Z^{\prime }$-mass and the weak phase in
models with (a) $\protect\beta =-1/\protect\sqrt{3}$ and (b) $-\protect\sqrt{%
3}$ for different ansatz, where $R_{M}$ is the rotation matrix in the Matsuda ansatz, $R_{MN}$ in the
Matsuda-Nishiura ansatz, $R_{F}$ in the Fritzsch ansatz, and $R_{CKM}$ assuming 
$\left|\widetilde{B}_{L}^{sb}\right|=\left|V_{tb}V_{ts}^{\ast}\right|$.}}}
\label{fig-bs}
\end{figure}

The left-handed coupling in Eq. (\ref{chiral-mass}) contains non-diagonal
components, which induce mixing between the neutral $Z^{\prime}$ boson and quarks from different families. This will produce new physics
contributions to the mass difference in neutral meson systems as for example
in Kaons $K^{0}-\overline{K^{0}},$ Bottom $B_{d}^{0}-\overline{B_{d}^{0}}$ \
and Bottom-strange $B_{s}^{0}-\overline{B_{s}^{0}}$ mesons, each one induced by the $s-%
\overline{d},$ $d-\overline{b}$ and $s-\overline{b}$ transition,
respectively. In particular, we take the most recent data of the $B_{s}^{0}$ difference mass given by Eq. (\ref{massdiff}) in order
to constraint new physics induced by the $Z^{\prime}$ interaction. The ratio between the experimental
value and the SM prediction in Eq. (\ref{massdiff}) is \cite{mesones}

\begin{equation}
\frac{\Delta m_{s}^{\exp }}{\Delta m_{s}^{ME}}=\left\vert 1+3.57\times
10^{5}e^{2i\phi _{L}^{sb}}\left( \frac{M_{Z}}{M_{Z^{\prime }}}\widetilde{B}%
_{L}^{sb}\right) ^{2}\right\vert =0.894\pm 0.243,  \label{meson-mezcla}
\end{equation}

\noindent with $\widetilde{B}_{L}^{sb}$ the $s\overline{b}$ component of $\widetilde{B}%
_{L}^{D}$ defined by Eq. (\ref{chiral-mass}), and $\phi _{L}^{sb}$ the weak
phase. The above data constrain the values of the $Z^{\prime}$
mass and the weak phase assuming different ansatz in the texture of the mass
matrices of the quarks, as discussed in Sec. 3. For the rotation matrix $%
R_{D}$ in the down sector, we consider the Fritzsch ansatz (%
$R_{F}$) in Eq. (\ref{Fritzsch-biunitary}), the Matsuda-Nishihura ansatz ($R_{MN}$) in Eq. (\ref{Matsuda2-Biunitary}), and the
Matsuda ansatz ($R_{M}$) in Eq. (\ref{Matsuda-Biunitary}). In order to achieve a complete comparison, we also consider the flavor-changing contribution assuming that $\left|\widetilde{B}_{L}^{sb}\right|=\left|V_{tb}V_{ts}^{\ast}\right|$, with $V_{tb(ts)}$ the $t-b(t-s)$ component of the CKM matrix, where we use the values $\left|V_{tb(ts)}\right|=0.77(4.06\times10^{-4})$ \cite{data-particle}. We use the notation $R_{CKM}$ for this last case. Figs. \ref{fig-bs} show plots of
the contours at $1 \sigma$ C.L for both (a) $\beta =-1/\sqrt{3}$ and (b) $%
\beta =-\sqrt{3}$ models, and for each ansatz of the rotation matrix, where we use the following data at the $Z$ scale

\begin{eqnarray}
M_{Z} &=&91.1876\pm 0.0021\text{ GeV};\text{\quad }S_{W}^{2}=0.23113\pm
0.00033;  \notag \\
m_{u}(M_{Z}) &=&1.38\;\text{MeV};\qquad m_{d}(M_{Z})=3.05\;\text{MeV};\qquad m_{c}(M_{Z})=0.626\;\text{GeV}; 
\notag \\
m_{s}(M_{Z}) &=&58.04\;\text{MeV;}\qquad m_{b}(M_{Z}) =2.89\;\text{GeV};\qquad m_{t}(M_{Z})=171.8\;\text{GeV}.
\label{weak-parameters-Z}
\end{eqnarray}

\noindent The regions below each curve correspond to excluded points in the $%
M_{Z^{\prime}}-\phi _{L}^{sb}$ plane. We can also see excluded regions in
the center of each curve, as shown in the plots. The minimum values for $%
M_{Z^{\prime}}$ are found when $\phi _{L}^{sb}=\pi /2.$ In particular, the
ansatz $R_{CKM}$ induces the lowest bounds with values $M_{Z^{\prime
}}\approx 1$ TeV, while the Matsuda ansatz $R_{M}$ leads high values with bounds from $M_{Z^{\prime }}\approx 10$ TeV for $\beta =-1/\sqrt{3}$ models, and $%
M_{Z^{\prime }}\approx 30$ TeV for $\beta =-\sqrt{3}.$ Although the flavor-changing
contribution is very sensitive to the rotation matrix, we find points that overlap
regions from different ansatz in the curves of the plots. For example, in fig. \ref{fig-bs}-(a),
we find the same value $M_{Z^{\prime }} = 10$ TeV for $R_{M}, R_{MN}$ and $%
R_{F}$ if $\phi _{L}^{sb}=5 \pi /12, 5 \pi /24$ and $\pi /12$, respectively.

The differences exhibited by each curve in the above figures arise from the size of the mixing components of the couplings $\widetilde{B}_{L}^{D}$ for each ansatz. In tab. \ref{tab:left-handed-couplings}, we compare the non-diagonal components of the left-handed coupling in the down sector. We also compare the FCNC contribution for the $u\overline{c}$ component in the up sector, as shown in the last line from Tab. \ref{tab:left-handed-couplings}. First of all, we observe that the maximum mixing resides in the $b-s$ sector, which is about one order of magnitude bigger than other flavor-changing transitions, like for example in $K^{0}-\overline{K^{0}},$ $%
B_{d}^{0}-\overline{B_{d}^{0}}$ and $D^{0}-\overline{D^{0}}$ systems, which induce the flavor changing
transitions $s \leftrightarrow d$, $b \leftrightarrow d$, and $u \leftrightarrow c$, respectively. Secondly, the Matsuda ansatz yields the biggest couplings, so that in order to control the low energy limits exhibit by Eq. (\ref{meson-mezcla}), it is necessary to impose stronger restrictions to the new physics contribution induced by this ansatz, such as seen in figs. \ref{fig-bs}. 

\begin{table}[tbp]
\begin{center}
\begin{tabular}{ccccccccc}
\hline
&  &  &  &$\left\vert \widetilde{B}_{L}^{Q} \right\vert (\times 10^{-2})$ &  &  &  &\\ 
\hline\hline
\multicolumn{1}{c|}{$D^{\prime }\overline{D}$} & $R_{M}$ &  & \multicolumn{1}{|c}{$R_{MN}$} &  & \multicolumn{1}{|c}{$R_{F}$} &  & \multicolumn{1}{|c}{$R_{CKM}$} & 
 \\ \hline\hline
$\beta $ & \multicolumn{1}{|c}{$-1/\sqrt{3}$} & \multicolumn{1}{|c}{$-\sqrt{3}$} & \multicolumn{1}{|c}{$-1/\sqrt{3}$} & \multicolumn{1}{|c}{$-\sqrt{3}$} & \multicolumn{1}{|c}{$-1/\sqrt{3}$} & \multicolumn{1}{|c}{$-\sqrt{3}$} & \multicolumn{1}{|c}{$-1/\sqrt{3}$} & \multicolumn{1}{|c}{$-\sqrt{3}$} \\ \hline\hline
$d\overline{s}$ & \multicolumn{1}{|c}{$5.8$} & 
\multicolumn{1}{|c}{$17.6$} & \multicolumn{1}{|c}{$0.6$} & 
\multicolumn{1}{|c}{$1.9$} & \multicolumn{1}{|c}{$0$}  & \multicolumn{1}{|c}{$0$} & \multicolumn{1}{|c}{$0.01$} & \multicolumn{1}{|c}{$0.04$}  \\ \hline
$d\overline{b}$ & \multicolumn{1}{|c}{$6.0$} & 
\multicolumn{1}{|c}{$18$} & \multicolumn{1}{|c}{$2.7$} & 
\multicolumn{1}{|c}{$8.2$} & \multicolumn{1}{|c}{$1.6$} & \multicolumn{1}{|c}{$3.7$} & \multicolumn{1}{|c}{$0.4$} & \multicolumn{1}{|c}{$1.2$} \\ \hline
$s\overline{b}$ & \multicolumn{1}{|c}{$26.0$} & \multicolumn{1}{|c}{$78.7$} & 
\multicolumn{1}{|c}{$11.9$} & \multicolumn{1}{|c}{$36.0$} & \multicolumn{1}{|c}{$7.4$} & \multicolumn{1}{|c}{$22.5$} & \multicolumn{1}{|c}{$2.2$} & \multicolumn{1}{|c}{$6.8$} \\ \hline
$u\overline{c}$ & \multicolumn{1}{|c}{$0.076$} & \multicolumn{1}{|c}{$0.23$} & 
\multicolumn{1}{|c}{$0.11$} & \multicolumn{1}{|c}{$0.33$} & \multicolumn{1}{|c}{$2.25$} & \multicolumn{1}{|c}{$5.3$} & \multicolumn{1}{|c}{$--$} & \multicolumn{1}{|c}{$--$} \\ \hline
\end{tabular}%
\end{center}
\caption{\textsf{{\protect\small Magnitudes of the Left-Handed couplings for different ansatz of the rotation matrix. We consider 331 models with $\protect\beta =-1/%
\protect\sqrt{3},$ and $\protect\beta =-%
\protect\sqrt{3}.$}}}
\label{tab:left-handed-couplings}
\end{table}

In addition to the low energy differences shown by the plots in the above figures, the different sizes of the coupling $\widetilde{B}_{L}^{D}$ leads to different predictions of the decay width of the $Z^{\prime}$ boson into quarks. In particular, the flavor-changing width can be written as   

\begin{equation}
\Gamma _{Z^{\prime }\rightarrow \overline{q}q^{\prime }}=\frac{%
g_{L}^{2}M_{Z^{\prime }}}{16\pi C_{W}^{2}}\left[ \left( \widetilde{g%
}_{v}^{qq^{\prime }}\right) ^{2}+\left( \widetilde{g}_{a}^{qq^{%
\prime }}\right) ^{2}\right]=\frac{%
g_{L}^{2}M_{Z^{\prime }}}{8\pi C_{W}^{2}}\left[ \left( \widetilde{B%
}_{L}^{qq^{\prime }}\right) ^{2}\right],  \label{Z'-width-FCNC}
\end{equation}

From Tab. \ref{tab:left-handed-couplings} it is evident that the main source of flavor-changing decay is $Z^{\prime} \rightarrow b\overline{s}$, with a decay probability of about $80\%$ bigger than others flavor-changing decays. A detailed study of FCNC decays is carried out in ref. \cite{z2-decay} in the Matsuda ansatz. Other ans\"atze, as the ones considered here, will yield lower values in the width.

\section{Conclusions}

In the framework of the 3-3-1 models, we have described the contribution to the mass difference $\Delta M_{s}$ in $B_{s}$ meson systems. These models behave as a purely left-handed neutral flavor-changing model. Using the recent experimental data and the SM one-loop prediction of $\Delta M_{s}$, we found bounds for the mass of the $Z^{\prime }$
boson in the Foot-Long-Truan model (FLT) and the
Pisano-Pleitez-Frampton model (PPF). The lowest values of $M_{Z^{\prime }}$ are found when the weak angle associated to the $b-s-Z^{\prime}$ coupling is $\phi _{L}^{sb}=\pi /2.$ By assuming four different ans\"atze in the texture of the mass
matrices of the quarks, we obtained plots of the allowed regions in the $%
M_{Z^{\prime}}-\phi _{L}^{sb}$ plane. We considered the Fritzsch ansatz (%
$R_{F}$), the Matsuda-Nishihura ansatz ($R_{MN}$), and the
Matsuda ansatz ($R_{M}$) for the rotation matrix $%
R_{D}$ in the down sector. We also assumed another alternative, where $\left|\widetilde{B}_{L}^{sb}\right|=\left|V_{tb}V_{ts}^{\ast}\right|$. Lower bounds from $M_{Z^{\prime }}$$\approx1$ TeV to $\approx10$ TeV in the FLT model, and from $M_{Z^{\prime }}$$\approx2$ TeV to $\approx30$ TeV in the PPF model are found for each ansatz of the rotation mass matrix. Since the Matsuda ansatz leads to the biggest size of the left-handed couplings as shown by tab. \ref{tab:left-handed-couplings}, this ansatz exhibit stronger low energy limits than the other ans\"atze. Also, the Matsuda texture yields a bigger probability of flavor-changing decay for the $Z^{\prime }$ boson than the other ans\"atze.

\vspace{0.3cm}

This work was partially supported by Fundaci\'{o}n Banco de la Rep\'{u}blica and by ALFA-EC funds through the HELEN programme. F. Ochoa would like to thank F. Schrempp for helpful discussions and hospitality at DESY, where part of this work was developed in the framework of the HELEN programme.

\end{document}